\documentclass[journal]{vgtc}                     

\onlineid{0}
\PassOptionsToPackage{dvipsnames,svgnames}{xcolor}
\usepackage{xcolor}
\usepackage{tcolorbox}

\newcommand{\myparagraph}[1]{\noindent{\textbf{#1}}}

\usepackage{fixltx2e}

\vgtccategory{Research}

\newcommand{\review}[1]{\textcolor{black}{#1}}
\newcommand{\prop}[1]{\textsf{#1}}

\vgtcpapertype{please specify}

\title{StreetWeave: A Declarative Grammar for Street-Overlaid Visualization of Multivariate Data}

\author{%
  Sanjana Srabanti, G. Elisabeta Marai, and Fabio Miranda
}

\authorfooter{
  \item
  	Sanjana Srabanti, G. Elisabeta Marai, and Fabio Miranda are with the University of Illinois Chicago.
  	E-mail: \{ssraba2,gmarai,fabiom\}@uic.edu
}

\abstract{%
The visualization and analysis of street and pedestrian networks are important to various domain experts, including urban planners, climate researchers, and health experts. 
This has led to the development of new techniques for street and pedestrian network visualization, expanding possibilities for effective data presentation and interpretation.
Despite their increasing adoption, there is no established design framework to guide the creation of these visualizations while addressing the diverse requirements of various domains. 
When exploring a feature of interest, domain experts often need to transform, integrate, and visualize a combination of thematic data (e.g., demographic, socioeconomic, pollution) and physical data (e.g., zip codes, street networks), often spanning multiple spatial and temporal scales. 
This not only complicates the process of visual data exploration and system implementation for developers but also creates significant entry barriers for experts who lack a background in programming.
With this in mind, in this paper, we reviewed 45 studies utilizing street-overlaid visualizations to understand how they are applied in practice. 
Through qualitative coding of these visualizations, we analyzed three key aspects of street and pedestrian network visualization usage: their analytical purposes, the visualization approaches employed, and the data sources used in their creation. 
Building on this design space, we introduce StreetWeave, a declarative grammar for designing custom visualizations of multivariate spatial network data across multiple resolutions.
We demonstrate how StreetWeave can be used to create various street-overlaid visualizations, enabling effective exploration and analysis of spatial data. StreetWeave is available at \href{https://urbantk.org/streetweave}{urbantk.org/streetweave}.
}

\keywords{Urban visual analytics, design space, street-overlaid visualization, visualization grammar.}

\teaser{
  \centering
  \includegraphics[width=1\linewidth]{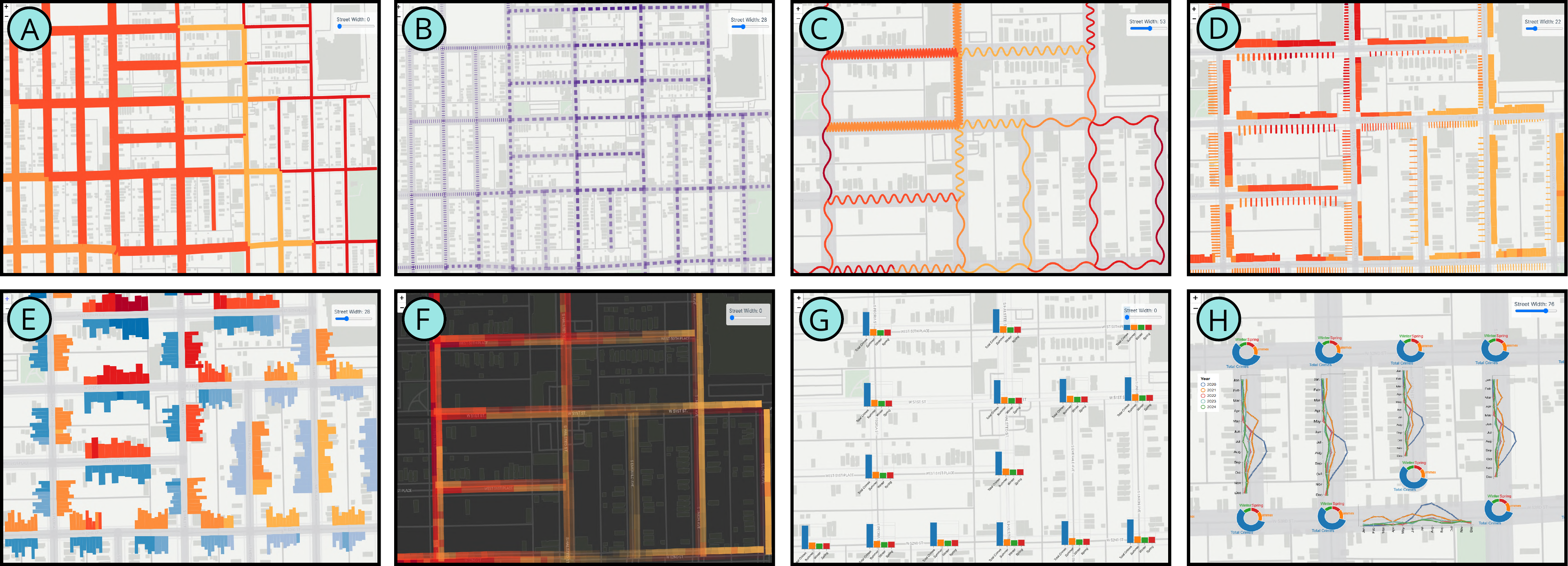}
  \caption{Street and pedestrian network visualizations with  StreetWeave. (A) A multivariate line map, with attributes like color and line width. (B) \& (C) A pattern-based encoding approach for street overlays, utilizing dashed lines, variable opacities, and wave patterns to represent data attributes. (D) \& (E) A bristle map and a line map that communicate data through variations in width, height, color, and opacity, with an additional alignment dimension integrating multiple encodings. (F) A multi-dimensional grid, where each cell encodes a distinct data value, and the background map color can be customized. (G) \& (H) Charts applied directly to street intersections and segments, with an orientation dimension that allows for adjustments of plot orientation.}
  \label{fig:teaser}
}

\newcommand{\hide}[1]{}

\graphicspath{{figs/}{figures/}{pictures/}{images/}{./}} 

\usepackage{tabu}                      
\usepackage{booktabs}                  
\usepackage{lipsum}                    
\usepackage{mwe}                       

\usepackage{mathptmx}                  

\usepackage{xcolor}

\usepackage{amsmath}
\usepackage{amsfonts}
\usepackage{amssymb}
\usepackage{algorithm}
\usepackage{algpseudocode}

\definecolor{Why}{HTML}{78C679}
\definecolor{How}{HTML}{FE9929}
\definecolor{What}{HTML}{807DBA}
\definecolor{purple}{HTML}{a73fd3}

\newcommand{\rectangle}[2]{\tikz[baseline=(Y.base)] \node[draw=none, fill=#1, text=white, rounded corners=2pt, minimum height=1em, inner sep=1pt,align=center,font=\sffamily\scriptsize,anchor=center,text height=1.2ex,text depth=0ex] (Y) {#2};}
\newcommand{\expert}[1]{\rectangle{purple}{#1}}

\begin{document}



\maketitle

\section{Introduction}

Street-overlaid visualization has a rich history that dates back to 1854, when John Snow created one of the most iconic examples: a map of the cholera outbreak in London.
By overlaying cholera cases on a street map of London, he could communicate the source of the outbreak\cite{nollenburg2007geographic}. 
Since then, street-overlaid visualizations have become popular across various domains, including climate and environmental sciences~\cite{peters2022assessing, hu2024street, yang2024thermal, guo2024mechanism, fu2024urban, rahman2024unveiling}, urban planning~\cite{woo2008airscope, steed2003science}, and urban accessibility studies~\cite{hara2014scalable, tajgardoon2015simulating}, as they provide in-depth insights into multiple factors. 
%
%
At the same time, visualizing street and pedestrian networks is challenging due to their complex structures, vast datasets, and diverse analytical requirements. Meaningful visualization outcomes often depend on interdisciplinary collaborations involving experts from computer science, urban planning, transportation engineering, architecture, geography, and public health~\cite{mariano2020designing}. These collaborations typically produce specialized workflows for analyzing and visualizing data spanning various spatial and temporal scales.
While visualizations of street and pedestrian networks have supported diverse analytical tasks across domains, including identifying pollution hotspots~\cite{steed2003science} or highlighting accessibility concerns~\cite{steed2003science, woo2008airscope}, the research community remains fragmented~\cite{yap2022free}. Many projects arise from collaborations between computer scientists and domain experts focused on specific problems, resulting in one-off solutions that are difficult to adapt or reuse across contexts or regions. 
Creating them demands skills in visualization, graphics programming, and data integration, raising entry barriers for those without technical expertise. While recent progress in visualization toolkits and grammars has lowered some technical barriers~\cite{moreira2023urban}, there is still a lack of a structured and dedicated approach tailored specifically for visualizing street and pedestrian networks.

Given the significance of street and pedestrian network visualizations, understanding their design space is \review{necessary}. Prior work has explored design spaces for networks, geospatial networks, and map visualizations. For instance, Schöttler \review{et al.'s} survey reviewed visualization and interaction techniques for geospatial networks, particularly in contexts like social networks, and trade and migration~\cite{schottler2021visualizing}, while Hogräfer \review{et al.} surveyed map-like visualization techniques~\cite{hografer2020state}. 
However, there has not yet been a dedicated effort to create a design space specifically for street and pedestrian network visualizations.
Visualizing data on streets poses several challenges, including overplotting, limited space for information encoding, and the risk of obscuring important spatial details~\cite{sun2014embedding, sun2016embedding}. Additionally, aggregating data by geographic regions leads to a loss of spatial specificity, while simultaneously encoding multiple attributes on streets can result in visual clutter~\cite{kim2013bristle}. This limitation highlights the need for more structured guidance on effectively embedding multivariate data directly onto streets, ensuring a clear, integrated, and meaningful understanding of spatial information.
To address this need, we conducted a qualitative study analyzing \review{45} existing visualizations overlaid on street maps. Through a systematic review of these works, we examined the analytical objectives that motivated the use of street-overlaid visualizations, the visual components employed to construct them, and the data types used to drive these visual encodings. We abstracted common tasks such as identifying specific segments of interest (e.g., high-crime or congested streets), comparing spatial or temporal conditions (e.g., weekday vs. weekend traffic, neighborhood accessibility), and analyzing patterns (e.g., uncertainty in travel, shifting volumes, regional movement, or congestion spread).  Additionally, we analyzed how designers used visual channels, marks, and compositions to represent information directly on street networks. Finally, we studied the forms of data, ranging from raw observations to derived metrics, that informed these visualizations. This analysis provides a broad understanding of how street-overlaid visualizations are constructed and used across domains.
Our design space serves as a systematic and actionable framework for developing visualizations tailored to diverse analytical needs for street and pedestrian networks. By leveraging this structure, StreetWeave supports task-specific visualization design, ensuring that users can effectively explore and interpret spatial data within their areas of interest. 

Building on these insights, we introduce StreetWeave, a declarative grammar designed to simplify the creation of street and pedestrian network visualizations. StreetWeave lowers entry barriers by reducing the need for deep technical knowledge. Specifically, it supports easy prototyping, clear evaluation of visualization designs, and flexible adaptation across various applications and geographic contexts. The system is reusable and extensible, supporting visualization development across a range of scenarios and domains. \review{StreetWeave's grammar-based approach eliminates the need for low-level code manipulation, enabling users to focus on what to visualize rather than how to implement it. This makes the system especially beneficial for our primary target users, urban planners, transportation engineers, public health analysts, climate researchers, and other domain experts, who often work with spatial and thematic data but may not have extensive programming experience.}
%
%
\review{StreetWeave aims to bridge the gap between complex visualization techniques and practical usability. It supports reuse, cross-domain applications, and offers a simplified approach to embedding information directly within street networks. Unlike abstract network analysis tools, StreetWeave treats streets as structural scaffolds for organizing data while preserving geographic context. 
}
By lowering entry barriers, StreetWeave allows users to integrate diverse datasets directly onto streets and pedestrian pathways, enhancing the way we analyze and interact with urban spatial data. We demonstrate how StreetWeave enables the creation of diverse street and pedestrian network visualizations, facilitating \review{clear} exploration and in-depth analysis of spatial data across various applications.
Our key contributions are:

\begin{itemize}[noitemsep, topsep=0pt]
     \item We propose a design space for street and pedestrian network visualizations, based on a review of 45 papers, organizing insights by analytical tasks, visualization techniques, and data types.
     \item We introduce StreetWeave, a high-level visualization grammar designed specifically for street and pedestrian network visualizations, enabling users to design interactive visualizations with minimal technical effort. StreetWeave is available at \mbox{\href{https://urbantk.org/streetweave}{urbantk.org/streetweave}}.
     \item We report on a set of use cases inspired by previous works demonstrating StreetWeave's ease of use, flexibility, and reproducibility.
 \end{itemize}

\section{Related Work}

We begin by reviewing prior work on the design space of geospatial visualizations, then examine street-overlaid visualizations, and finally explore related work on visualization grammars and authoring tools. 

\subsection{Design spaces for geospatial visualizations}

Data has always been \review{important} for understanding and navigating spaces\cite{wieczorek2009geographic}. From early hand-drawn maps to modern visual analytics systems~\cite{miranda2024state, ferreira2024assessing}, advancements in data collection and analysis have enabled the creation of highly accurate and insightful maps\cite{zhang2024mapcraft}.
This progress has highlighted the need for structured design spaces for map and geospatial network visualizations.
Hogräfer et al.~\cite{hografer2020state} proposed a design space for map-like visualizations by categorizing existing techniques from two perspectives: imitation and schematization. They focused on understanding how abstract data could resemble traditional maps (imitation) or how maps could be made more abstract for specific purposes (schematization). Each of these perspectives is divided into four categories based on the affected visual elements: point, line, area, and field. Their work provides a comprehensive taxonomy for map-like visualizations by grouping similar techniques within each category. Similarly, Zhang et al.~\cite{zhang2024mapcraft} analyzed over 100 geo-infographics to establish a design space for geo-infographic creation, focusing on four main dimensions: basic map representations, data encoding channels, label design and placement, and highlighting techniques for key information. They even created a web-based authoring tool, MapCraft, to make geo-infographic creation accessible to a wider audience. Additionally, Schöttler et al.~\cite{schottler2021visualizing} conducted a survey exploring visualization and interaction techniques for geospatial networks, covering multiple fields to create a design space. This work categorizes visualization techniques based on geography representation, network depiction, composition, and interactivity. Despite the existence of design spaces for various geospatial visualizations, there is still no dedicated design space for street and pedestrian network visualizations that specifically guide the creation of these types of visual representations.

\subsection{Street-overlaid visualizations}

Street-overlaid visualizations have emerged as an approach for exploring and analyzing multivariate spatiotemporal data, effectively integrating geographical, temporal, and attribute-based information~\cite{kim2013bristle, sun2014embedding, sun2016embedding}. 
\review{These visualizations display data directly on the geometry of street or pedestrian networks, using streets as the visual anchor to encode additional data attributes.}
As the volume and complexity of spatial data continue to grow, these visualizations enable users to uncover patterns, trends, and anomalies that are otherwise hard to detect while preserving the spatial structure of the network.
%
%
Street-overlaid visualizations play a key role in several types of analyses, such as public transportation routes, route choice behavior, trajectory analysis, and overall traffic patterns\cite{fawcett2000adaptive, wang2013visual, wang2014visual, zeng2014visualizing, wang2014visual2, lu2015visual, zeng2016visualizing, al2016semantictraj, andrienko2016leveraging, pi2019visual, he2019interactive, andrienko2019visual, weng2020towards}. 
%
Moreover, several new methods have been developed to further enhance street-overlaid visualizations. Techniques such as occlusion-free route zooming\cite{sun2014embedding}, street enlargement\cite{sun2016embedding}, and aggregation methods for multilevel data analysis\cite{deng2022multilevel} have all contributed to making these visualizations more effective and easier to interpret.
In addition to transportation analysis, street-level visualizations play a significant role in crime analysis\cite{garcia2021cripav, garcia2020mirante}. By overlaying crime incident data on street maps, analysts can quickly identify crime hotspots, monitor changes over time, and understand spatial crime patterns. This helps in developing targeted interventions and policies for improving public safety. 
Similarly, in urban planning, visualizing urban data directly at the street level provides critical insights into infrastructure needs, population density, and accessibility, facilitating better planning and decision-making processes\cite{tajgardoon2015simulating, huang2015trajgraph, wilson2018quantifying, levashev2016modern, feng2020topology}. Street-overlaid visualizations allow urban planners to evaluate scenarios, identify areas needing improvement, and create plans that enhance connectivity, accessibility, and quality of life. 

\subsection{Visualization grammars and authoring tools}

Visualization grammars and authoring tools have \review{become widely adopted} approaches \review{for simplifying} the creation of interactive visualizations. Generally, these tools are grouped into three main categories: template-based tools, low-level visualization libraries, and high-level visualization grammars~\cite{ferreira2024assessing}.
The template-based visualization tools typically utilize graphical interfaces, allowing users to create visualizations through drag-and-drop interactions without requiring programming knowledge~\cite{wongsuphasawat2017voyager}. 
Despite their ease of use, these tools offer limited flexibility and customization, making it difficult to handle the complex analytical tasks associated with street and pedestrian network visualizations. 
On the other hand, low-level visualization libraries, such as D3~\cite{bostock2011d3}, provide considerable freedom in creating custom visualizations. However, their use demands programming expertise, limiting accessibility to non-technical users. Additionally, these libraries often lack built-in capabilities for easily integrating multiple data layers and interactions. 
High-level visualization grammars represent a middle ground, offering a flexible yet accessible way to create visualizations through declarative specifications. Tools like Vega~\cite{Satyanarayan7192704}, Vega-Lite~\cite{Satyanarayan7539624}, the Urban Toolkit (UTK)~\cite{moreira2023urban}, and similar grammar-based frameworks allow users to define visualizations through concise and declarative specifications. Users can quickly experiment with different visualization types, encoding strategies, and interaction methods without writing extensive code. Although powerful, these grammars typically lack specialized support for \review{directly} integrating and interacting with multiple thematic and physical layers specifically associated with street networks.

Inspired by these existing grammars and authoring tools, our work on StreetWeave aims to overcome these limitations. StreetWeave introduces a high-level visualization grammar specifically designed for creating street and pedestrian network visualizations. \review{It allows users to create complex visualizations using simple, declarative rules, eliminating the need for low-level programming.} StreetWeave also supports the integration of multiple thematic and spatial data layers directly onto street networks, offering both flexibility and usability for visualization and domain experts alike.

\section{StreetWeave's Design Goals}

The design of StreetWeave follows a structured approach informed by prior research on street and pedestrian network visualizations~\cite{kim2013bristle, sun2014embedding, wang2014visual, sun2016embedding, andrienko2016leveraging}, a review of \review{45} existing studies, and our analysis of key challenges. Our goal was to streamline the process of designing, prototyping, and sharing interactive visualizations of street and pedestrian networks. This comprehensive review shaped the development of StreetWeave, which unfolded in three main phases.
In the first phase, we systematically analyzed the existing literature to identify limitations and tradeoffs in previous visualization tools and practices, with the results presented in Section~\ref{sec:space}. Next, we reflected on our personal experiences as visualization researchers and tool builders, focusing on ways in which a unified framework could enhance current methodologies. This reflection allowed us to define clear, detailed design objectives and capabilities for StreetWeave. 
Finally, during the framework's implementation phase, we continuously evaluated whether our established design goals were effectively met. 
\review{Next, we present StreetWeave's design goals, inspired by our own UTK~\cite{moreira2023urban}.}
We refer to these goals throughout the subsequent sections, particularly when discussing StreetWeave's grammar (Section~\ref{sec:grammar}) and usage scenarios (Section~\ref{sec:usage}). We also reflect on how well these goals were achieved in Section~\ref{sec:reflection}.

\myparagraph{Ease of use and minimal technical barriers.} Creating street and pedestrian network visualizations often requires programming skill, restricting their accessibility to non-technical domain experts. To address this issue, \review{StreetWeave must provide a straightforward declarative grammar that simplifies complex technical tasks such as data loading, rendering, and interaction design.} By reducing technical complexity, StreetWeave must allow users to focus on visual analysis and exploration, instead of low-level implementation details, making advanced visualization accessible even to those without specialized programming experience.

\myparagraph{Rapid prototyping and flexibility.} Visual analytics workflows are inherently exploratory, often requiring users to iteratively test and refine visualization ideas. Traditional tools can hinder this process due to the extensive programming effort involved in adapting visualizations for new questions, datasets, or geographical contexts. StreetWeave must support rapid prototyping by enabling users to easily experiment with different visualization techniques, data encodings, and interactions within the same declarative framework. Additionally, StreetWeave must allow flexible adaptation across regions, scales, and datasets without requiring extensive reimplementation, supporting broader use by domain experts beyond the tool's original development context.

\myparagraph{Extensibility.} Beyond rapid iteration, street and pedestrian network analysis requires applying custom visual designs across diverse spatial scales, aggregation levels, and thematic layers. Insights often develop through shifting analytical perspectives and evolving hypotheses, which can be hindered by rigid tooling. StreetWeave must enable users to define and extend custom visualizations that align with varied analytical workflows, including advanced operations like data aggregation and comparative spatial analysis. \review{Additionally, StreetWeave must support embedding Vega-Lite specifications within its grammar, leveraging a wealth of easy-to-use visualizations from the Vega community to extend its capabilities.}

\myparagraph{Reproducibility.} Reproducing street and pedestrian network visualizations remains challenging. Many tools are closed-source and depend on complex, custom features and specialized libraries, making replication difficult even when data is available~\cite{ferreira2024assessing}. StreetWeave must address reproducibility through two strategies. First, it must use standard input data formats, allowing users to easily package and share datasets used in visualizations. Second, it must treat visualization configurations as structured, declarative specifications, so reproducing a visualization requires only sharing the high-level specification and corresponding datasets, enabling straightforward verification and reuse.

\section{StreetWeave's Design Space}
\label{sec:space}

To develop a clear understanding of how street and pedestrian network visualizations are created and used, we first conducted a systematic survey of existing works. We compiled a diverse collection of 45 papers, focused on street-overlaid visualizations across various domains to identify common usage patterns and practices.


\subsection{Selection process}
To gain a comprehensive understanding of street and pedestrian network visualizations, we began by searching for papers that featured street-overlaid visualizations. Initially, we focused on proceedings and collections from major venues where visualization research is typically published, such as IEEE Transactions on Visualization and Computer Graphics, Computer Graphics Forum, ACM Conference on Human Factors in Computing Systems, and IEEE PacificVis. For each of these venues, we manually scanned the proceedings and retrieved papers based on their titles. If a title contained keywords such as `street', `trajectory', `route', or `road', we included those papers in our collection. We then manually examined each paper to confirm if it contained street-overlaid visualizations, discarding any that did not. Through this process, we initially gathered 25 papers from these venues.
To expand our collection, we then conducted additional searches on Google Scholar using the same keywords. Through this method, we identified a substantial number of additional papers from various venues. We followed the same manual scanning process to verify whether these papers included street-overlaid visualizations, ultimately adding 20 more relevant papers to our collection.

\subsection{Data coding}

\review{After compiling the 45 papers featuring street-overlaid visualizations, we conducted a structured thematic coding process to analyze how these visualizations were constructed and for what purposes. The first author individually reviewed each paper to verify inclusion criteria and then systematically coded each figure using a framework informed by prior geospatial visualization literature \cite{hografer2020state, schottler2021visualizing, zhang2024mapcraft}. The coding focused on three key dimensions: (1) analytical tasks (e.g., identify, compare, pattern analysis), (2) visual encodings (e.g., color, opacity, width, orientation), and (3) data types (e.g., direct observations vs. derived metrics). Each paper could include multiple visualization instances, and all were carefully annotated to record how these dimensions were used.
Although coding was performed by a single researcher, detailed notes and criteria were maintained to ensure consistency across the dataset. The frequency of appearance for each visual encoding, task category, and data type was quantified and is reported in the corresponding sections of the paper (Sections~\ref{sec:analytical_task},~\ref{sec:visual_component}, and~\ref{sec:data_abstraction}), where we break down the number of papers exhibiting each element. We acknowledge that including a coding protocol with multiple coders would further strengthen reliability and plan to incorporate that in future extensions of this work.}

\subsection{Design framework}

To understand how these visualizations are constructed and used in practice, we structured the design space around common analytical tasks, visualization techniques, and data types observed across a broad set of research papers.

\subsubsection{Analytical task abstraction}
\label{sec:analytical_task}
Understanding the analytical goals behind street-overlaid visualizations is essential for designing tools that effectively support user tasks.
To identify these goals, we reviewed 45 research papers spanning domains such as urban planning, transportation, crime, and climate studies, extracting common analytical tasks supported by street-overlaid visualizations.
In some cases, this extraction was challenging, particularly in papers centered on visualization techniques or algorithms without explicitly stated analytical objectives.
Nevertheless, our review revealed several recurring analytical tasks across the surveyed studies, specifically 
\textbf{identify}, \textbf{compare}, and \textbf{pattern analysis} tasks. \review{While these high-level tasks serve as an initial abstraction, StreetWeave explicitly targets concrete, realistic analytical scenarios informed by our qualitative analysis of existing street-overlaid visualizations.}


\textbf{Identify} tasks focus on locating specific spatial areas, such as street segments, intersections, or sidewalk regions, that satisfy a known condition. 
These goal-driven tasks appeared in 93\% of the reviewed papers. Examples include identifying roads with the highest nighttime crime rates~\cite{garcia2020mirante, garcia2021cripav}, locating areas with high bicycle usage~\cite{he2019interactive}, pinpointing congested street segments~\cite{wang2014visual, pi2019visual}, or highlighting sidewalks that are inaccessible to people with disabilities~\cite{tajgardoon2015simulating}.
\review{In urban planning, a targeted example is detecting sidewalk segments with significant accessibility issues (e.g., missing or inadequate curb ramps) to help planners prioritize interventions~\cite{li2024never}.}
\textbf{Compare} tasks involve evaluating differences or similarities across spatial regions, such as streets, intersections, or neighborhoods.
These tasks facilitate the assessment of spatial variation across the street network and were present in 88\% of the reviewed papers. Examples include comparing traffic flow between streets~\cite{wang2013visual, andrienko2016leveraging, pi2019visual} or assessing accessibility scores across different road segments~\cite{tajgardoon2015simulating}.
\review{A specific example is comparing sidewalk conditions between adjacent neighborhoods to guide equitable resource allocation~\cite{li2024never, tajgardoon2015simulating}.}
\textbf{Pattern analysis} tasks center on detecting broader spatial or spatiotemporal trends and relationships across street networks. This category appeared in 62\% of the reviewed papers.
Examples include analyzing uncertainty in traffic behavior, monitoring changes in traffic volume~\cite{wang2014visual, pi2019visual}, identifying mobility or population flow patterns~\cite{zeng2014visualizing}, evaluating connectivity within street networks~\cite{andrienko2019visual}, \review{or crime hotspots and clusters of service requests along streets to inform targeted safety or maintenance initiatives~\cite{garcia2020mirante, garcia2021cripav}.}

\subsubsection{Visual component abstraction}
\label{sec:visual_component}

This dimension examines the visual elements used in street-overlaid visualizations to align encoding choices with analytical tasks.  We categorize these elements into \textbf{visual channels} (e.g., color, size, opacity), \textbf{marks} (e.g., lines, points, glyphs), and \textbf{compositions} (e.g., juxtaposition, superimposition).
Color hue was the most frequently used channel (78\% of the papers), allowing clear differentiation of features~\cite{lu2015visual, cornel2015visualization, tajgardoon2015simulating, wilson2018quantifying, deng2022multilevel}, while color intensity (33\%) conveyed variations in data values~\cite{kim2013bristle, weng2020towards, garcia2020mirante} supporting identify, compare, and pattern analysis tasks.
Opacity (7\%) indicated data density or confidence through transparency~\cite{kim2013bristle}. Geometric properties such as length, width, and height further supported multivariate spatial data representation, while additional channels like size, density, dashed lines, and blur encoded complementary attributes.
For marks, lines (51\%) were commonly used, often combined with color and opacity for multivariate encoding. Points and circles conveyed data at specific locations, while waves and curved lines occasionally depicted uncertainty or directional flows. Glyphs provided compact representations of multivariate and spatiotemporal data, supporting complex pattern analysis and comparison~\cite{borgo2013glyph, huang2015trajgraph}. Labels in text or symbol form aided direct identification and interpretation~\cite{huang2015trajgraph, garcia2020mirante}.
In terms of composition, we observed that juxtaposition and superimposition were the dominant strategies in street-overlaid visualizations. Juxtaposed views (38\% of the papers) placed visualizations side-by-side for coordinated comparison~\cite{sun2014embedding, sun2016embedding, weng2020towards}, while superimposed views (used in all papers) layered visual elements directly on the street network~\cite{kim2013bristle, pi2019visual, deng2022multilevel}, supporting interpretation while risking clutter when encoding multiple variables.

\subsubsection{Data abstraction}
\label{sec:data_abstraction}
Our review shows that street-overlaid visualizations primarily use data that can be categorized into two broad types: \textbf{direct observation} and \textbf{derived metrics}, both requiring geographic references (e.g., latitude/longitude or segment identifiers) to align data with street networks.
\textbf{Direct observations} refer to raw data collected without further transformation or spatial computation. Examples include GPS traces capturing pedestrian or vehicle trajectories~\cite{he2019interactive, pi2019visual}, sensor readings recording traffic speed~\cite{fawcett2000adaptive}, and manual audits of sidewalk conditions~\cite{tajgardoon2015simulating}. These values can be mapped directly onto the street network based on their geographic coordinates or segment associations.
\textbf{Derived metrics} are generated through processing, aggregation, or spatial analysis over raw data. These include summary statistics such as the average traffic volume on a street segment over a week~\cite{wang2013visual, pi2019visual}, the flood vulnerability recorded at intersections~\cite{cornel2015visualization}, or aggregated counts of crime incidents along streets~\cite{garcia2020mirante, garcia2021cripav}. 
Derived metrics can also involve complex spatial computations, such as calculating accessibility scores~\cite{tajgardoon2015simulating, wilson2018quantifying}.


\section{StreetWeave's Grammar}
\label{sec:grammar}

Based on the design goals and insights from our systematic review, we introduce StreetWeave, a declarative visualization grammar tailored for street and pedestrian network visualizations. StreetWeave provides clear, structured rules for specifying visualizations, covering spatial units (e.g., area, segment, node), multiple data layers (thematic and physical data integrated via spatial relations), and diverse visual encoding options. Spatial units serve as the grammar's foundation, with all other specifications defined relative to each visualization layer.
In the following sections, we detail the grammar, organizing it into three parts: (1) general grammar specification, (2) data layer definitions (including thematic and physical data, spatial relations, and analytical operations), and (3) data plotting specifications. For formal presentation, we adopt the notation of Ren et al.~\cite{10.1109/TVCG.2018.2865158}, where := denotes assignment, | indicates ``or'', ? marks optional elements, and + represents one or more elements.

\begin{figure*}[!t]
\centering
\includegraphics[width=1\linewidth]{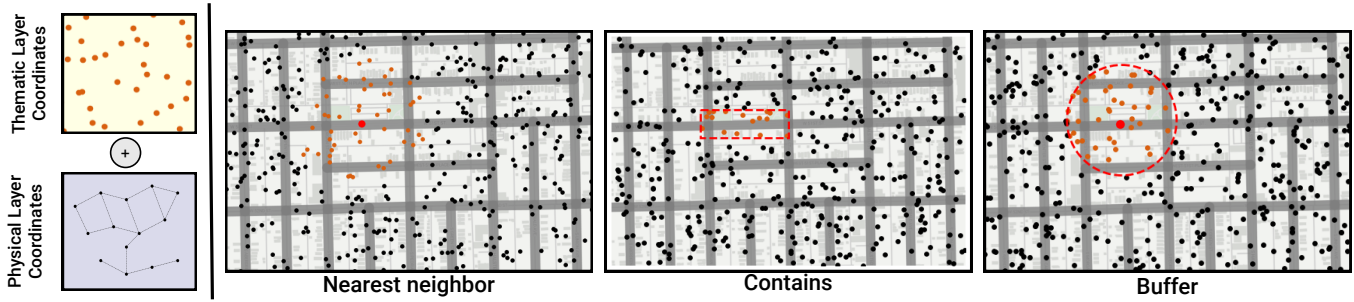}
\caption{Integrating thematic and physical layers through spatial relations. StreetWeave integrates thematic data with the physical street network by applying spatial relations at each street segment or intersection. The nearest neighbor relation selects the 50 closest data points to a segment or intersection, while the contains relation includes only those within a segment's or intersection's boundaries. The buffer relation gathers points within a user-defined radius from a segment's or intersection's midpoint. Aggregation operations like mean or sum can be applied to these points for analysis.}
\label{fig:Relation}
\end{figure*}

\subsection{General specification}
\label{sec:gen_grammar}

To structure the StreetWeave visualization grammar, we begin by defining a clear general specification. Each layer in the grammar comprises three key components: a map, a unit (Section~\ref{sec:unit_grammar}), and data layer aggregations (Section~\ref{sec:data_grammar}).
\review{All properties include recommended default settings to guide users through the visualization process while minimizing the need for extensive manual adjustments, particularly for those with limited technical expertise.}


\begin{tcolorbox}[colback=gray!3, colframe=gray!30]
\begin{description}[leftmargin=0cm, labelsep=0.0em, itemsep=0.0em, style=sameline, font={\bfseries\sffamily},before={\sffamily}]
    \item[StreetWeave] := (map$^+$, unit$^+$, data$^+$, relation$^?$, query$^?$)
    \item[map] := streetColor$^?$ , streetWidth$^?$, background$^?$
    \item[background] := `light' | `dark' 
\end{description}
\end{tcolorbox}


The main configuration defines the foundational visual and interactive settings for presenting street and pedestrian network visualizations.
It specifies \prop{map}, \prop{unit}, \prop{data}, \prop{relation}, and \prop{query}.
Additional map attributes allow customization of the visualization's appearance, such as selecting the \prop{background} style as light or dark, setting \prop{streetColor} to any desired color, and specifying \prop{streetWidth} to control line thickness.

\subsection{Unit and visual encodings specification}
\label{sec:unit_grammar}

In StreetWeave's grammar, the concept of a \prop{unit} defines both the spatial granularity for data aggregation and analysis, and the corresponding visualization specification.


\begin{tcolorbox}[colback=gray!3, colframe=gray!30]
\begin{description}[leftmargin=0cm, labelsep=0.0em, itemsep=0.0em, style=sameline, font={\bfseries\sffamily},before={\sffamily}]
    \item[unit] := type, density$^?$, method, opacity$^?$, color$^?$, dash$^?$, squiggle$^?$, width$^?$, height$^?$, chart$^?$, rows$^?$, columns$^?$, orientation$^?$, alignment$^?$, zoom$^?$
    \item[type] := `segment' | `node' | `point'
    \item[method] := `line' | `rect' | `matrix'
    \item[chart] := vega\_spec
    \item[orientation] := `parallel' | `perpendicular'
    \item[alignment] := `left' | `center' | `right'
    \item[zoom] := [ minZoom , maxZoom ]
\end{description}
\end{tcolorbox}

The StreetWeave grammar supports flexible spatial resolutions through the \prop{type} property, which specifies units as \textit{segment}, \textit{node}, or \textit{point}. A \textit{segment} represents individual street segments and can be subdivided using the \prop{density} property, which accepts a fixed value or a thematic variable for finer granularity, such as visualizing sidewalk severity (Figure~\ref{fig:Example2}). 
\review{This subdivision enables users to control the spatial resolution of their analysis, allowing denser sampling in areas of interest while maintaining coarser aggregation elsewhere. By mapping \prop{density} to a data attribute, users can simultaneously encode an additional variable, supporting multivariate analysis through combinations of \prop{density} and other visual properties (e.g., color, width).}
A \textit{node} corresponds to intersections, enabling aggregation and visualization of data at street junctions (Figure~\ref{fig:Example3}). A \textit{point} allows the placement of point-based thematic data, such as event or sensor locations, directly onto the network.
\review{StreetWeave also supports scale-dependent rendering through the \prop{zoom} property, which specifies the zoom levels at which a unit will be displayed. The \prop{zoom} property is defined as a range, \texttt{[minZoom, maxZoom]}, so that users can control the visibility of units depending on the current map zoom level. This feature enables the creation of multi-scale visualizations, where detailed information can be shown at higher zoom levels and clutter is reduced when viewing the network at broader scales.}

StreetWeave's grammar enables embedding visual encodings directly on street networks using two approaches. The first utilizes graphical methods specified by \prop{method} (e.g., \textit{line}, \textit{rect}, \textit{matrix}), which can be customized with attributes such as \prop{color}, \prop{opacity}, \prop{width}, and \prop{height}. Additional stylistic properties like \prop{dash} and \prop{squiggle} allow visual differentiation, aiding in the representation of uncertainty or categorical distinctions (Figure~\ref{fig:teaser} (B, C)). 
\review{Importantly, these properties can either be assigned fixed values or mapped dynamically to data attributes, supporting the representation of quantitative or categorical information directly within the visualization.}
The second approach leverages the \prop{chart} property, allowing users to embed Vega-Lite specifications directly into StreetWeave visualizations. This integration provides access to a wide range of pre-defined visualization types, facilitating the inclusion of complex, multivariate encodings while maintaining \prop{alignment} and \prop{orientation} on the street network. 
%
\review{The \prop{orientation} property further extends expressive capacity by allowing users to render visual elements either \textit{parallel} or \textit{perpendicular} to the street network. For example, rendering a perpendicular bar chart along a street segment while encoding color parallel to the segment can help users interpret traffic volume alongside congestion levels without visual overlap.}

\subsection{Data and spatial relations specification}
\label{sec:data_grammar}

StreetWeave's grammar enables users to load, visualize, integrate, and apply analytical operations directly on data layers, classified into two distinct categories: thematic layers and physical layers. 
\review{This separation, inspired by our prior work on UTK~\cite{moreira2023urban}, facilitates clear organization of spatial data for analysis and visualization.}
Thematic layers represent data attributes measured at specific locations, such as crime incidents, pollution levels, or noise complaints. Typically, thematic data consists of discrete points, each with spatial coordinates and associated values. In contrast, physical layers depict structural or environmental features, including road networks, pedestrian networks, and intersections. \review{They define the geometric reference on which visualizations are overlaid, giving spatial context for interpreting thematic data.}

\begin{tcolorbox}[colback=gray!3, colframe=gray!30]
\begin{description}[leftmargin=0cm, labelsep=0.0em, itemsep=0.0em, style=sameline, font={\bfseries\sffamily},before={\sffamily}]
    \item[data] := physical$^?$ , thematic$^?$
    \item[physical] := path
    \item[thematic] := path , latColumn$^?$ , lonColumn$^?$
    \item[relation] := spatial , value , type\textsubscript{relation}
    \item[spatial] := `buffer' | `nn' | `contains'
    \item[type\textsubscript{relation}] := `sum' | `mean' | `min' | `max'
\end{description}
\end{tcolorbox}



In addition, StreetWeave computes the bearing angle of street segments to capture their orientation, aiding in the analysis of movement patterns and directional dependencies within the street network.
The grammar supports spatial operations to integrate thematic and physical layers effectively (Figure~\ref{fig:Relation}). By defining a spatial relation (e.g., buffer with a specified radius, contains, or nearest neighbor), thematic data can be linked to physical geometries, enabling spatially grounded analyses.
When multiple thematic data points correspond to a single physical geometry, the grammar supports aggregation operations (e.g., mean, sum, max, min) to summarize these values. For example, pedestrian counts or noise levels recorded at discrete locations can be aggregated onto street segments or intersections based on the defined spatial relation. In doing so, StreetWeave allows analyses such as evaluating accessibility around intersections or summarizing pollution levels along street segments within a consistent spatial framework.


\begin{figure}[!t]
\centering
\includegraphics[width=1\linewidth]{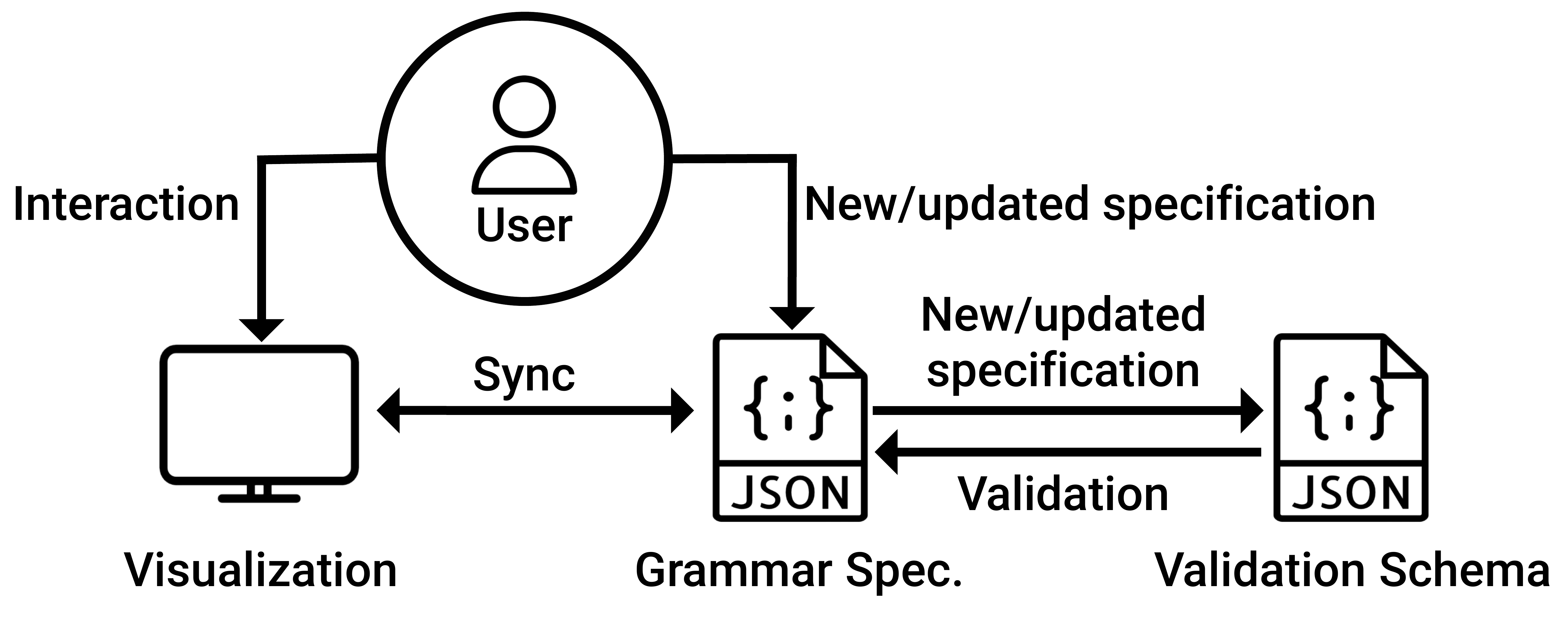}
\caption{Overview of StreetWeave's system architecture and execution pipeline. User specifications are processed, interpreted, and rendered to support interactive network visualizations.}
\label{fig:system_architecture}
\end{figure}

\subsection{Query specification}
\label{sec:que_grammar}

With StreetWeave, users can selectively modify styling (e.g., width) of segments using two optional query parameters: a geographic address and a radius around that address.


\begin{tcolorbox}[colback=gray!3, colframe=gray!30]
\begin{description}[leftmargin=0cm, labelsep=0.0em, itemsep=0.0em, style=sameline, font={\bfseries\sffamily},before={\sffamily}]
    \item[query] := address$^?$ , radius$^?$
\end{description}
\end{tcolorbox}

By specifying an \prop{address} and a \prop{radius}, users can define geographic areas within which width modifications will apply, supporting localized adjustments without affecting the entire network. For example, users can emphasize streets within a specific neighborhood to highlight areas with high pedestrian activity, event impact zones, or regions undergoing construction.

\subsection{Implementation}
StreetWeave is implemented as a grammar-based toolkit using JavaScript.
\review{Users specify visualizations by authoring a JSON-based specification that follows the grammar previously introduced.}
\review{This specification is parsed and validated against StreetWeave's JSON schema to ensure syntactic correctness and to guide users during the authoring process. Users explicitly define the layers, visual encodings, and operations required for their visualization, which are then rendered directly on the street or pedestrian network within the interface.}
Input data includes physical layers provided in JSON format and thematic layers supplied as CSV files.
StreetWeave's interface consists of two primary components: a text editor and a map view panel. The text editor allows users to input and edit their grammar specifications, which are validated using a JSON schema to ensure correctness and provide guidance. Users reference their physical and thematic data layers directly within the specification. Once specified, visualizations are rendered in the map view panel, allowing users to explore and refine designs iteratively.
\review{Figure~\ref{fig:system_architecture} presents a high-level overview of StreetWeave's system architecture.}
StreetWeave uses React.js for interface management, Leaflet.js for interactive map rendering, and D3.js for custom SVG-based overlays and data-driven visual encodings. 
By integrating these libraries, StreetWeave supports smooth interaction capabilities, including panning and zooming, while maintaining efficient rendering performance even with complex, multivariate visualizations.
To support extensibility and user experimentation, StreetWeave allows the embedding of Vega-Lite specifications within its grammar.
\review{An interpreter reads the user's JSON-based visualization specification and dynamically generates corresponding SVG elements, which are then appended to the map. Because StreetWeave operates on a layer-based map architecture, it can leverage any mapping platform that supports layer management.}

\section{Usage Scenarios}
\label{sec:usage}

To demonstrate StreetWeave's capabilities, flexibility, and ease of use, we present a series of scenarios grounded in real-world data challenges. These examples illustrate how StreetWeave can integrate diverse datasets to generate detailed street-overlaid visualizations that support in-depth analysis and interpretation.


\begin{figure*}[!t]
\centering
\includegraphics[width=2\columnwidth]{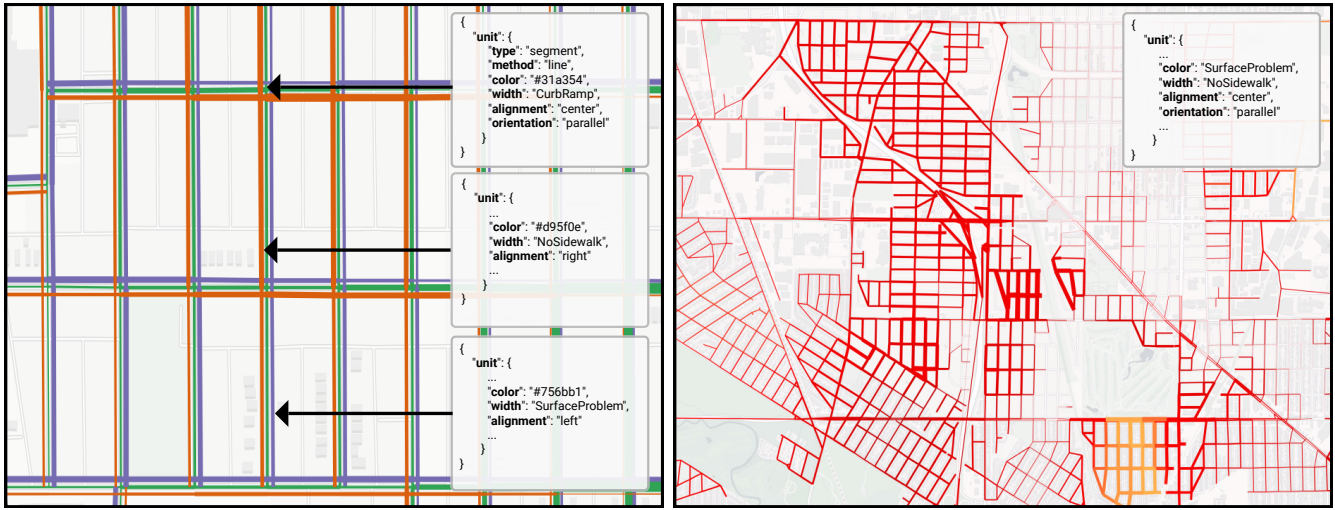}
\caption{Visualizing the sidewalk accessibility severity. \review{Left: A multi-line visualization maps sidewalk accessibility attributes onto street segments, using line width to encode data values across three overlaid layers: a center-aligned green line encodes curb ramp severity, a right-aligned brown line represents missing sidewalks, and a left-aligned purple line shows surface problem severity.} Right: A multivariate encoding combines color and width to represent multiple data attributes simultaneously. The corresponding StreetWeave specifications demonstrate the grammar-based approach that configures and adapts these multivariate encodings.}
\label{fig:Example1}
\end{figure*}

\subsection{Scenario 1: Analyzing sidewalk accessibility data}
\label{sec:Example1}

To illustrate StreetWeave's capabilities and flexibility, we present an analysis scenario using the Project Sidewalk~\cite{saha_project_2019} dataset for Chicago, which contains geo-located data on sidewalk accessibility issues, such as severity of crosswalk and curb ramp problems, missing curb ramps, absent sidewalks, obstacles, and surface issues. 
In this example, we visualize the distribution and severity of these accessibility issues across Chicago's street network, enabling simultaneous exploration of multiple concerns.
Using StreetWeave's grammar, we define a visualization where each line segment encodes different thematic attributes through \prop{width}, allowing users to analyze various conditions in a single integrated view.
We begin by loading the physical layer data (geometric information for street segments) and the thematic layer data (sidewalk severity ratings in CSV format). Using StreetWeave's grammar within the text editor, we specify spatial relations between thematic and physical layers, applying a \textit{buffer} relation with a 10-meter radius and a  \textit{mean} aggregation type to summarize sidewalk severity onto street segments.
For visualization, we select \textit{segment} as the unit type to display data at the street level. The center-aligned line encodes curb ramp severity through its width. A left-aligned purple line encodes the severity of the surface problem using width. Additionally, a right-aligned line represents missing sidewalks, also using width for encoding severity.

After specifying these configurations, we apply the specification, and the map view panel instantly updates to display the multi-encoded visualization (Figure~\ref{fig:Example1} (left)). Users can interact with the map by zooming and panning to identify problematic street segments and prioritize areas for infrastructure improvements. Because sidewalk accessibility data is highly multivariate, StreetWeave's ability to handle multiple data attributes within a single view is particularly valuable, enabling users to analyze patterns and identify critical areas based on several factors.

\subsection{Scenario 2: Detailed sidewalk accessibility analysis}
\label{sec:Example2}

\begin{figure*}[!t]
\centering
\includegraphics[width=1\linewidth]{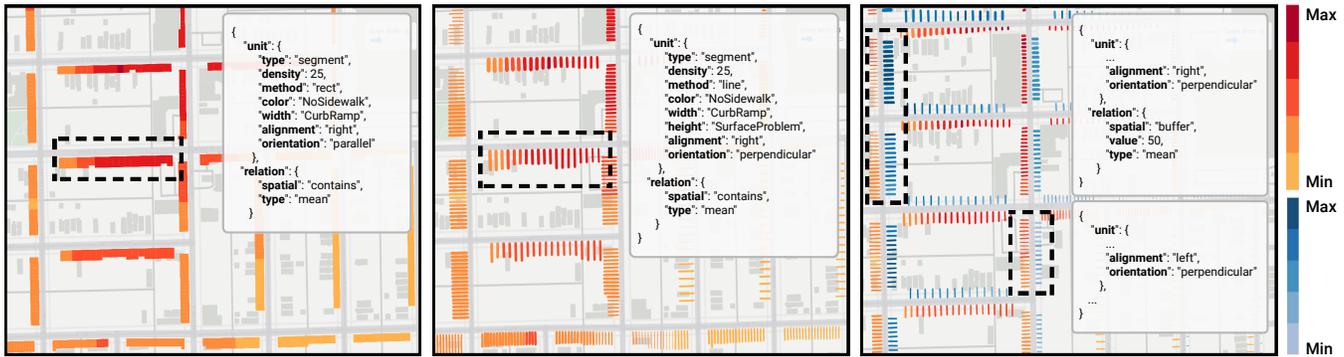}
\caption{Fine-grained street segment analysis. Left: A line map displays sidewalk accessibility severity using variations in color and width. Middle: A bristle map incorporates an additional height dimension to more accurately highlight problematic areas, as seen in the highlighted regions. Right: Dual bristle maps, aligned to the left and right, visualize multivariate data side-by-side, providing another perspective on localized accessibility issues.}
\label{fig:Example2}
\end{figure*}

In the previous example, we demonstrated how StreetWeave can identify problematic street segments across Chicago by visualizing sidewalk accessibility data directly on the streets. However, marking an entire street segment as problematic may not provide the granularity needed for targeted urban improvements. To pinpoint which portions of a segment exhibit accessibility challenges, a finer-grained analysis is necessary. 
For this example, we again utilize the Project Sidewalk dataset but we now focus on sub-segment granularity by subdividing each street segment into smaller sections. This enables spatial aggregation and visualization on each sub-segment, providing more accurate and detailed insights into sidewalk accessibility conditions.
To visualize this enhanced detail, we first create a line map that encodes multiple accessibility conditions using \prop{color} and \prop{width}, with the visual encoding oriented \textit{parallel} to the street (Figure~\ref{fig:Example2} (left)). By simply changing the \prop{orientation} to \textit{perpendicular}, \prop{method} to \textit{line}, and adding the \prop{height} dimension, we produce a bristle-style visualization inspired by Bristle Maps~\cite{kim2013bristle} (Figure~\ref{fig:Example2} (middle)). In this variant, each perpendicular ``bristle'' extending from the street represents a sub-segment, with its visual attributes directly reflecting sidewalk accessibility data. Users can encode multiple variables simultaneously in different combinations, such as color and height or color and width, providing clear visual cues that highlight different aspects of the data. This configuration allows users to observe detailed variations in accessibility conditions within a single street segment, which may be too subtle to detect in a parallel line map.

This rapid reconfiguration is achieved by modifying a few attributes in StreetWeave's grammar. Users can explore and compare multiple visualization setups without additional effort. StreetWeave also supports the simultaneous display of multiple visualizations on the same map, such as displaying different bristle maps on either side of the street, enabling direct comparisons and deeper spatial analysis. In Figure~\ref{fig:Example2} (right), two layers of bristle maps are shown side by side, illustrating how different accessibility conditions can vary along the same street segment, with some conditions showing low severity while others exhibit higher severity.

\subsection{Scenario 3: Exploring crime and 311 service data}
\label{sec:Example3}

\begin{figure}[!t]
\centering
\includegraphics[width=1\linewidth]{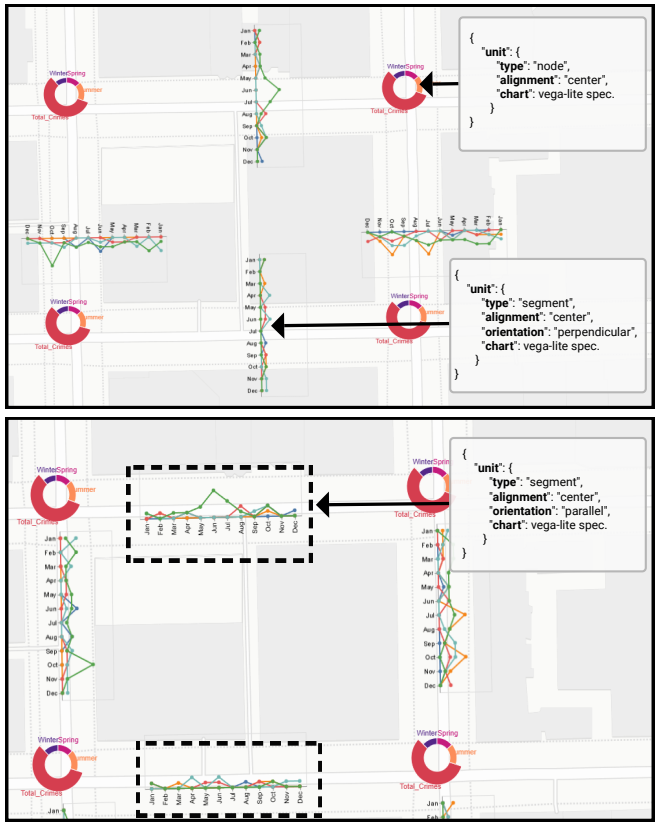}
\caption{Visualizing crime patterns and 311 service requests. Top: Charts are applied to street intersections (radial charts) and street segments (line charts) using perpendicular orientation, enabling multi-layered representation. Bottom: Charts are rendered parallel to the streets for clearer presentation. This reveals noticeable contrasts in service call distributions, even between adjacent streets (as highlighted by the rectangle).}
\vspace{-0.4cm}
\label{fig:Example3}
\end{figure}

In addition to sidewalk accessibility, understanding patterns in urban crime and city service requests is essential for creating safer and more responsive urban environments. For this scenario, we demonstrate how StreetWeave can simultaneously visualize and analyze two distinct yet interrelated datasets: crime incidents and 311 service requests in Chicago, showcasing StreetWeave's flexibility and support for iterative visualization design.
We begin by visualizing crime incident data aggregated at street intersections (nodes). Leveraging StreetWeave's integration with Vega-Lite, we embed compact charts directly within the visualization, using \textit{contains} as the spatial relation and \textit{mean} as the aggregation type to summarize crime data at each intersection. This approach highlights the distribution of crime across intersections while demonstrating StreetWeave's extensibility through seamless incorporation of external visualization libraries.
Next, we analyze 311 service requests from 2020 to 2024, plotting them along street segments. We again use the \textit{contains} spatial relation, now with the \textit{sum} aggregation type, to represent the frequency of service requests along each segment. Initially, we applied a \textit{perpendicular} orientation for these encodings (Figure~\ref{fig:Example3} (top)). However, upon reviewing the visualization, we found that switching the \prop{orientation} to \textit{parallel} (Figure~\ref{fig:Example3} (bottom)) improved readability by aligning the encodings with street paths, facilitating easier trend tracing.
%
As shown in Figure~\ref{fig:Example3} (bottom), clear differences in service request distributions are visible even on adjacent streets (highlighted by the rectangle), revealing varying service needs within small geographic areas and underscoring StreetWeave's capacity for detailed spatial analysis.
\review{The visualizations in this scenario serve as illustrative examples of StreetWeave's grammar, demonstrating that it is not limited to these encodings. The underlying grammar remains extensible, supporting the incorporation of additional types or custom visual components as needed for diverse analytical tasks.}

This iterative design process highlights StreetWeave's strengths in rapid prototyping, enabling users to experiment with different visualization parameters and immediately assess their effectiveness. By facilitating the simultaneous visualization of crime and service request data, StreetWeave helps users explore potential relationships or divergences between crime hotspots and areas with high service demand. These insights can assist urban planners and city officials in prioritizing safety and service improvements effectively.

\section{Discussion}

In this section, we first discuss feedback gathered from domain experts, then compare our contributions with existing visualization tools, and finally reflect on how StreetWeave's design goals were met.

\subsection{Expert feedback}

\review{To evaluate StreetWeave's utility and alignment with real-world needs, we conducted semi-structured interviews with five experts working extensively in the urban domain. None of the participants are authors of this paper. Participants included one expert with a PhD in Urban Systems \expert{E1}, one with a PhD in Computer Science \expert{E2}, one with a PhD in Civil Engineering \expert{E3}, one PhD student in Computer Science \expert{E4}, and one PhD student in Urban Planning \expert{E5}. Quotes from participants have been lightly edited for clarity.}

\review{Ease of use and workflow simplification emerged as a consistent theme, with experts appreciating how StreetWeave can reduce the complexity of creating multivariate, street-overlaid visualizations compared to D3, GeoPandas, or multi-step pipelines. \expert{E3} noted that she ``has used D3, and it is complicated,'' adding that StreetWeave ``is easy (...) it is good that you do not need to think about the code beneath it.'' Similarly, \expert{E4} noted that she ``never felt comfortable with D3.'' While \expert{E3} acknowledged that ``multilayers can be done in GeoPandas,'' she explained that it required ``creating multiple maps,'' whereas ``StreetWeave can easily do this.'' \expert{E2} emphasized that ``while D3 is valuable for heavy customization, it requires effort to create visualizations.'' She highlighted StreetWeave's usefulness for analyzing climate data, including temperature, rain, and wind, and visualizing heat waves and urban heat islands at street level. \expert{E4} highlighted that ``it normally takes a very long time to create street-overlaid visualizations, requiring multiple software packages,'' while StreetWeave ``provides a simple domain-specific language that even non-coders can use.''}
\review{
\expert{E5}, who had not used Vega-Lite before, found StreetWeave very interesting and expressed that he would have used it in previous projects had he known about it earlier. He appreciated the ability to ``set up a map once and then change the visualization quickly without extra coding,'' contrasting this with his usual workflow in GeoPandas, where ``you need to know what you want beforehand,'' whereas StreetWeave ``lets you quickly try out different design ideas to see what works.'' \expert{E5} noted that while Python felt intuitive for spatial work, StreetWeave's declarative commands offered faster iteration on visual styles, reducing the need to repeatedly adjust styles manually as he currently does in GeoPandas.
}

\review{\expert{E1} highlighted a promising new application for StreetWeave in the uncertainty analysis of computer vision models used for built environment audits, noting that StreetWeave ``could enable per-segment evaluation rather than more coarse block-level visualizations.'' However, \expert{E1} also pointed out a potential limitation, explaining that many models do not work at the segment level, but rather at a polygon level, such as models that extract sidewalk networks from satellite imagery. She noted that ``such cases would require additional computations to map the uncertainty to edges'' to fully leverage StreetWeave in these contexts.}

\review{Reproducibility was also noted as a major strength, with experts recognizing the value of sharing JSON specifications for replicable research workflows. \expert{E3} noted that ``you can share the JSON files and that definitely ensures the reproducibility, which is a big concern when you're doing research.'' \expert{E3} also emphasized that reproducibility extends beyond sharing static specifications, suggesting that coupling StreetWeave's outputs with clear data provenance could further strengthen transparency and facilitate validation across research projects.}

\review{The experts also suggested further extensions. \expert{E4} mentioned the option to embed functions within the grammar, ``giving advanced users more customization capabilities,'' as well as the ability to ``auto-detect coordinate reference systems to reduce friction for non-GIS users.'' Furthermore, experts noted the value of enabling StreetWeave to interoperate with standard tools like QGIS and ArcGIS.
\expert{E5} expressed interest in additional usability features, including the ability to add titles, legends, and axis labels, as well as options for exporting maps as PDF or image files for inclusion in reports and presentations.
}

\subsection{Comparison with existing tools}

In this section, we compare StreetWeave with widely used GIS tools like ArcGIS and QGIS. While powerful, these tools often overwhelm users with extensive operations and toolboxes requiring specialized knowledge~\cite{ziegler2023need}. 
They also lack direct support for effective street-overlaid visualizations, pushing users towards complex workflows or custom scripting.
In contrast, StreetWeave offers a concise, grammar-based approach tailored for street and pedestrian network visualizations. Its clear specification reduces complexity, enabling rapid prototyping and iteration without extensive technical expertise. 
Traditional GIS tools also struggle with reproducibility and easy sharing due to licensing constraints and complex dependencies.
StreetWeave addresses this by storing configurations as grammar specifications paired with JSON data, supporting collaboration and replication. Unlike existing GIS tools, it enables flexible, multivariate visualizations directly on street networks using diverse encodings, enhancing contextual interpretation and interactive analysis.


\review{In addition to GIS platforms, StreetWeave also differs significantly from conventional network visualization tools such as Gephi~\cite{bastian2009gephi}, Cytoscape~\cite{shannon2003cytoscape}, and Graphviz~\cite{ellson2002graphviz}, which are primarily designed to visualize abstract relational networks, such as social graphs, biological pathways, or citation networks, using algorithmic layouts like force-directed or radial positioning. These tools prioritize structural features like node centrality, clustering, and connectivity, often abstracting away geographic space entirely. In contrast, StreetWeave focuses on spatially grounded networks where the geometry of streets and intersections is essential to the analytical task. Rather than exploring topological relationships, StreetWeave supports multivariate spatial analysis by aligning thematic data directly with physical street geometry and enabling flexible visual encodings at multiple spatial resolutions. This makes StreetWeave more suitable for urban domains where tasks revolve around comparing infrastructure quality, accessibility, or socio-environmental indicators across real-world locations, rather than navigating purely relational graph structures.}


\subsection{Reflection on design goals}
\label{sec:reflection}

StreetWeave was designed to address the critical challenges of street and pedestrian network visualizations, emphasizing accessibility, flexibility, extensibility, and reproducibility. Through the examples demonstrated here, we have shown how StreetWeave meets each of these previous objectives.
In terms of \textbf{ease of use and minimal technical barriers}, Examples 1 and 2 illustrate the \review{accessible design} of the StreetWeave grammar. By simplifying the process of creating custom visualizations through an easy-to-understand, JSON-based grammar, StreetWeave lowers the entry barriers for non-technical users. The ability to define complex encodings, such as multivariate lines with varying orientations, not only enables nuanced and expressive visualization designs but also empowers visualization practitioners and domain experts to construct detailed analyses without requiring extensive programming knowledge, allowing them to focus directly on analytical insight rather than implementation.

The goal of \textbf{rapid prototyping and flexibility} is highlighted in Examples 2 and 3. By allowing quick alterations, such as changing the visualization orientations (from perpendicular to parallel) and method (from rect to line), StreetWeave facilitates iterative experimentation. Users can rapidly refine visualizations, testing various dimensions to identify the best presentations that effectively communicate complex data relationships. This capability shortens the iterative cycles common in exploratory visual analytics.

\textbf{Extensibility} is primarily addressed through the integration of external visualization capabilities, namely Vega-Lite, as shown in Example 3. This showcases how StreetWeave can incorporate an additional framework and embed external visual encodings without requiring fundamental changes to its underlying architecture. Users can \review{directly} extend the grammar to include diverse visualization types, providing the flexibility to adapt to evolving analytical demands.

Regarding \textbf{reproducibility}, StreetWeave emphasizes standardized data formats and declarative grammar specifications. Users can share their visual analyses by simply exchanging grammar specifications and datasets in JSON format, significantly enhancing the reproducibility and transparency of their workflows. This approach facilitates broader verification and adoption within both the visualization research community and among domain practitioners.

\section{Limitations, Conclusions, and Future Work}

\myparagraph{Limitations.}
\review{While StreetWeave demonstrates the potential of grammar-based street network visualization, it has a few limitations.
First, we have not yet conducted a formal user study to evaluate usability and effectiveness across users with diverse knowledge levels, domain expertise, and disciplinary backgrounds.
Second, we have not quantitatively assessed how comprehensively StreetWeave can reproduce visualization designs within the broader design space of urban street visualization, limiting our understanding of coverage and expressiveness.
Third, StreetWeave currently lacks a modular plugin system for custom visualization components outside of Vega-Lite, restricting more advanced and bespoke views.
Fourth, StreetWeave relies on SVG-based rendering, which can become slow when visualizing large street networks or several data layers, restricting performance in very complex scenarios.
Furthermore, the system's default values for encodings, colors, and sizing are based on practical experience rather than empirical studies, which may not align optimally with user preferences or perceptual best practices.
Finally, while the grammar simplifies reproducibility, integrating StreetWeave workflows with existing GIS or analysis pipelines still requires manual coordination, limiting end-to-end workflows.}

\myparagraph{Conclusions.}
In conclusion, we introduced StreetWeave, a declarative visualization grammar designed to streamline the creation, prototyping, and sharing of street and pedestrian network visualizations. Through a systematic review of 45 prior studies, we analyzed the analytical purposes, visualization techniques, and data types employed in street-overlaid visualizations. This analysis informed a structured design space that captures these insights, providing clear guidance to visualization designers across diverse domains.
To bring the design space into practice, we developed StreetWeave, a grammar-based specification specifically tailored for street and pedestrian network visualizations. StreetWeave reduces technical complexity, enabling users without extensive programming experience to design custom visualizations. The use cases presented demonstrate StreetWeave's ease of use, flexibility, and reproducibility in practical scenarios, including analyzing sidewalk accessibility and exploring crime and service request patterns.
\review{We see StreetWeave as a step towards bringing declarative grammars to urban-specific visualization challenges. By capturing urban domain requirements with a grammar, we hope to pave the way for broader adoption of visualization grammars in urban analytics.}

\myparagraph{Future work.}
\review{StreetWeave opens new research opportunities for advancing street network visualization. Much like how our evaluation study in Mota et al.~\cite{mota_comparison_2023} on 3D urban visualization catalyzed a series of papers expanding methods and applications, we hope StreetWeave will similarly spark future research in street-overlaid visualization.
Building on our prototype, we will conduct formal user studies across diverse domains and user expertise levels to evaluate the effectiveness of different street-overlaid visualizations. These studies will also help ground StreetWeave's default design and encoding choices in empirical evidence, aligning them with yet-to-be-uncovered perceptual best practices, guidelines, and user preferences to ensure effective communication for diverse urban analysis tasks. We also plan to quantitatively assess how comprehensively StreetWeave can recreate visualization designs within the broader urban street visualization design space, clarifying its coverage, expressiveness, and identifying gaps for further extension.
We will situate this work within our broader efforts on domain-specific visualization grammars, including the ongoing development of the UTK~\cite{moreira2023urban}, positioning StreetWeave as a reusable component in creating reproducible, urban-specific visualization tools.
We also envision extending StreetWeave to enable ``what-if'' analyses, allowing users to modify street structures, such as simulating widened roads or added pedestrian paths, and observe impacts on walkability, accessibility, and traffic flow. Achieving this will require deeper integration with simulation and predictive models, expanding StreetWeave's utility for urban planning, policy evaluation, and participatory design contexts.
Building on this, we see promising opportunities to develop grammars for broader scenario planning tasks, enabling experts to define and compare development or policy scenarios more systematically through linked visualization specifications.
From a technical standpoint, to address current performance limitations, we plan to port StreetWeave from SVG to other rendering systems such as WebGL and WebGPU, better providing scalable, interactive rendering for large networks, and supporting physical layer deformation, similar to UrbanRama~\cite{urbanrama}.
Additionally, we plan to integrate StreetWeave with the UTK and Curio frameworks~\cite{moreira2023urban, moreira_curio_2025}, enabling users to incorporate StreetWeave visualizations as modular components within broader urban data analysis workflows.
From a community standpoint, we also plan to create a ``StreetWeave Gallery'' for visualization specifications to be shared for reuse in other cities and projects.
}

\section*{Acknowledgments}
We thank the reviewers for their constructive feedback, the U.S. National Science Foundation (\#2320261, \#2330565, \#2411223), and the U.S. National Institutes of Health (R01CA258827, UG3TR004501).

\bibliographystyle{abbrv-doi-hyperref}

\bibliography{main}


\end{document}